\newcommand{\gev}{\ {\rm GeV}}
\newcommand{\mrad}{\ {\rm mrad}}

\newcommand{\zp}{Z.\ Phys.\ }
\newcommand{\pl}{Phys.\ Lett.\ }

\newcommand{\be}{\begin{equation}}
\newcommand{\ee}{\end{equation}}
\documentstyle[12pt]{article}
\newlength{\dinwidth}
\newlength{\dinmargin}
\begin{document}
\title{
\hfill {\large DESY~92--173}\\
\hfill {\large December 1992}\\
 \vspace{1 cm}
        {\bf Numerical study of radiative corrections in  the
         low $Q^2$ region  at HERA}
\author{
    K.Charchu\l a\thanks{On leave of absence from Institute of
    Experimental Physics, Warsaw University, Poland}
    \thanks{internet address: charchula@vxdesy.desy.de
        {\it or} krzych@fuw.edu.pl}
    \\[3pt]
    Deutsches Elektronen-Synchrotron, DESY, Germany  \\[10pt]
    J.Gajewski
    \thanks{internet address: gajewski@hozavx.fuw.edu.pl}
    \\[3pt]
    Institute of  Experimental Physics, Warsaw University, Poland\\[15pt]
        }
\date{ }
     }

\maketitle
%---------------
%\vspace{2 cm}

\begin{abstract}
A detailed  numerical study of  radiative corrections in
the low $Q^2$  region at the HERA $ep$ collider was performed.
The specific case of the total photoproduction cross section
measurement was taken as an example.
Two different programs, TERAD91 and HERACLES4.2, were used to get
an estimation of the size of radiative effects.
It was found that radiative corrections can be quite large
in some points of the space of leptonic $(x,y)$ variables.
However, after imposing experimentally feasible cuts on the radiated
photon and  the hadronic final  state  one gets   corrections at the
level  of a few per cent.
%\\
\end{abstract}
\setcounter{page}{1}
%\thispagestyle{empty}   % to suppress the page number on the first page
%\newpage
\vspace{1 cm}

\section{\bf Introduction}

The HERA $ep$ collider offers a very rich program of physics studies at
low $Q^2$ values \cite{sch92}. Since the cross section for
electron--proton
scattering is dominated by low $Q^2$ photon exchange,
the statistics of photoproduction events is large.
This fact allowed for a measurement of the total photoproduction
cross section at HERA \cite{zeu92,hhh92}
 already in the first running stage of HERA.

The  measured cross section is affected, in addition to various experimental
complications, also by effects coming from radiative corrections. In order to
determine the physically interesting Born part of  the      cross section one
usually  tries     to minimize the influence of radiative effects on  the
experimental data.

The treatment of radiative corrections (RC) to the photoproduction
process does not differ, in principle, from the corresponding analysis
for the more familiar     case of deep inelastic
scattering.  In fact, most results obtained for
the deep inelastic scattering case are also valid in the low $Q^2$
 region.
There exist, however, specific features of the photoproduction process
which should be carefully taken into account in  the study of
radiative effects.

The photoproduction events of  the process $ep\longrightarrow eX$
at HERA are characterized by small angles of the scattered
electron -- of the order of a few mrads. This translates into very low
values of virtuality of the exchanged photon -- down to a fraction of
the electron mass.
Because of the low $Q^2$ values involved  it is not possible, for
example,
to rely on the Quark Parton Model (QPM)
interpretation of the $ep$ process.
One needs also to keep the mass of the electron in the calculations.

In this note we  present a numerical study of RC in  the low $Q^2$
 region  at the HERA $ep$ collider  for  the specific case of
the measurement of the total photoproduction cross section with
the ZEUS detector \cite{zeu92}.
The analysis is based on two different programs:
TERAD91 \cite{ter92} and HERACLES4.2 \cite{her92}.
The first one (TERAD) uses a  semi--analytical approach
 and the other (HERACLES) is  based  on Monte--Carlo techniques.
 Both programs, which have been originally written for
the deep--inelastic range of kinematic variables, were recently
 modified  for   low $Q^2$   region \cite{bs92}.

\section{Some facts about radiative corrections}

Let us first recapitulate  the basic facts about radiative  corrections,
with emphasis on the specific region of our interest.

The cross section for the process
\be
\label{ep}
 e(p_e) + p(p_p)\longrightarrow e(p'_e) + X(p_X)
\ee
can be described most generally  in the language of structure
functions.
In the region of low $Q^2$ the interaction of the electron with the
proton is mediated only by the exchange of the photon. The lowest order
(Born) cross section  for the  neutral current (NC), parity conserving
interaction can be expressed in terms of  two structure
functions:
\begin{eqnarray}
\label{xborn}
\frac{d^2\sigma^{Born}(ep)}{dx_l dy_l} &=&
\frac{4\pi\alpha^2(s-M^2-m^2)}{Q^4_l}\ast\nonumber\\
&&
\left[
y^2_l \left( 1 - 2\frac{m^2}{Q^2_l}\right) xF^{em}_1(x_l,Q^2_l)
  + \left(1-y_l-\frac{M^2x_ly_l}{s-M^2-m^2}\right) F^{em}_2(x_l,Q^2_l)
\right]
\end{eqnarray}
where we retained explicitly the masses of the proton ($M$) and of the
electron ($m$). The leptonic variables $x_l, y_l$ and $Q^2_l$ are
defined in terms of  the scattered electron energy $E'_e$ and angle $\theta_e$
measured w.r.t. the electron beam:
\begin{eqnarray}
\label{qxy}
 Q^2_l &=& - (p_e-p'_e)^2 = x_ly_l(s-M^2-m^2)\\
       &=& -2m^2 + 2E_eE'_e \left(1 -
\frac{\sqrt{E_e^2-m^2}\sqrt{E_e'^2-m^2}\cos{\theta_e}}{E_eE'_e}\right)
\nonumber\\
y_l &=& \frac{p_p\cdot Q_l}{p_p\cdot p_e}
= 1 - \frac{E'_e}{E_e}\cos^2{\theta_e/2} +
{\cal O}\left(\frac{m^2}{E^2}\right)\\
x_l &=& \frac{Q^2_l}{2p_p\cdot Q_l}=\frac{Q^2_l}{y_l(s-M^2-m^2)}
\end{eqnarray}
 where $s=(p_e+p_p)^2$ and $E_e$ ($E_p$)   denotes  the energy of
 incoming electron (proton).
The  representation (\ref{xborn}) has a very important feature: it is
     valid both in  the deep inelastic region (large $Q^2 > M^2$) and
in the photoproduction region (small $Q^2 \sim m^2$). It is usually
 called model--independent \cite{bar92} in the sense that it
is derived from general invariance principles. In particular, it does
not rely on the QPM picture.

The measured cross section is, in general,  different from the Born
cross section. In addition to effects connected with the specific
experimental setup it includes also contributions coming from
higher order corrections to the one--photon exchange (usually called
radiative corrections). As a result one needs, for example,
to consider not only
the basic process (\ref{ep}) but also radiative processes with
additional photons:
\begin{eqnarray}
\label{inel}
{\rm   inelastic\  part:} &ep& \longrightarrow eX \gamma\\
\label{el}
{\rm elastic\ tail:}  &ep& \longrightarrow\ ep\gamma.
\end{eqnarray}
Radiative corrections spoil, in general, the simple relation
 (\ref{xborn}) between the cross section and the structure functions.
In order to disentangle the Born cross section from the measured cross
section one  usually tries to
reduce the  effects of RC by isolating radiative events which then
may be removed from the experimental data.

Various parts of       radiative corrections
 are usually classified according to the
Feynman diagrams they are originating from.
For the study of the photoproduction region only QED ($\equiv$  diagrams
with an additional real or virtual photon) and
leptonic ($\equiv$ related to the leptonic line of the Feynman diagram)
corrections are relevant. These are numerically largest   and
 can be treated in a model--independent  way, i.e., without any
assumptions based on the QPM  which is unreliable in the photoproduction
 region.
Our  analysis of     RC will be confined to
QED leptonic corrections to the electromagnetic (NC) interaction
calculated to order ${\cal O}(\alpha)$.

The size  of  RC strongly  depends on the definition of variables
with which they are studied (leptonic, hadronic, mixed, ...)
\cite{spi92}. The type of variables chosen is determined by the
possibility to access them in the  specific experimental situation.
In case of photoproduction it will be the leptonic variables.

The formula  for the radiative cross section expressed in
leptonic variables can be written as a sum of two parts:
\begin{eqnarray}
\label{xrad}
\frac{d^2\sigma^{\cal O(\alpha)}(x_l,y_l)}{dx_l dy_l} &=&
\left(1 + \delta^{vs}(x_l,y_l)\right)
\frac{d^2\sigma^{Born}(x_l,y_l)}{dx_l dy_l}\\
 &+&
 \int\int dx_h dy_h R(x_l,y_l; x_h, y_h)
\frac{d^2\sigma^{Born}(x_h,y_h)}{dx_h dy_h}\nonumber
\end{eqnarray}
where the hadronic variables $x_h,y_h$ are defined by the final
hadronic system.  In the presence of the photon emitted from the
electron line, hadronic variables differ from  the leptonic ones
and
\be
\label{hrange}
  x_h  \ge x_l, \qquad  y_h\le y_l.
\ee
The first part of (\ref{xrad}), proportional to $\delta^{vs}$,
 contains  contributions both coming from
virtual corrections and from the soft part of the real photon radiation
(soft bremsstrahlung). It corresponds to a "no observed photon"
situation. The second part accounts for the bremsstrahlung of hard
photons. The size of this contribution, in  contrary to the
first one, is sensitive to the  experimental conditions.
 It depends also on the structure functions, not only in a given
 $(x_l,y_l)$ bin, but in the range of $(x_h,y_h)$ defined by
(\ref{hrange}).
{}From the studies of deep inelastic scattering \cite{spi92} it is known
that large radiative corrections are mainly due  to the emission
of hard photons (the second part of (\ref{xrad})).
In the photoproduction case, the hard photon contribution
 vanishes in the limit of $Q^2\rightarrow 0$
\footnote{Thanks to H.Spiesberger for clarifying  this point.}.
However, in the kinematic region considered in this paper,
the influence of the hard photon radiation on the size of RC still
exists and is not negligible.

The size of radiative corrections can be significantly reduced
if events with hard photons are isolated. This can be done
in two ways. The straightforward  method consists of  identifying
hard emitted photons in the active part of a   detector.
They can be also separated indirectly by measuring
some characteristics of the hadronic final state. This follows from the
fact that cuts imposed on hadronic variables (for example, hadronic
invariant mass $W_h$) are equivalent to constraints on the radiated
photon kinematics.
In this study the influence of both methods on the size of RC will be
investigated.

\section{Programs for RC calculations}

The two programs used in our numerical study represent different
approaches to  RC calculations.

TERAD91 \cite{ter92} is a semi-analytical program based on formulas
 obtained  analytically,  including an analytical integration over the
photon   phase space. In addition,  one-  or twofold numerical
integration (depending on the set of variables used) is performed to
obtain the double differential cross section  $d^2\sigma/dx dy$, where
$(x,y)$ may   be leptonic, mixed or hadronic variables.
In case of QED leptonic corrections, one has an option to  use general
structure functions (not only those based on QPM).
For leptonic variables the phase space of the  hard photon
can be constrained in two ways: by cuts on the emitted photon energy
and angle, and by cuts on the hadronic final state
($Q^2_h$ and $W^2_h$).

Another approach was used in the HERACLES4.2 program \cite{her92}.
It is a Monte Carlo (MC) event generator, simulating the fully
differential  cross section. It means that besides calculating the cross
section  which includes  contributions from virtual and real radiative
corrections in a (predefined by user) kinematical region,
it also provides the 4--momenta of the final state particles in a
generated event. In particular, that of the radiated hard photon
is provided.
The MC approach allows to apply in a convenient way any phase space
restrictions. Also in this program it is possible  to use arbitrary
structure   functions  to describe  the $ep$ cross section.

As the size  of  RC is strongly related to experimental conditions,
the MC type of programs are better suited for these kind of
investigations.  On the other hand the semi-analytical approach can
serve  as an   independent check of results and also  allows some
experimental  constraints to be imposed.
For these reasons we used the above two programs in our analysis.

In both programs we used the ALLM prescription \cite{abr91}
for the structure functions, which provides the parametrization valid in
a wide range of $Q^2$ from the deep inelastic scattering region to the
$Q^2=0$ limit.

\section{A case study: leptonic QED corrections to
$\sigma_{tot}(\gamma p)$}

In our numerical analysis we will refer to the experimental setup
used in a recent measurement of the total photoproduction
cross section $\sigma_{tot}(\gamma p)$ with the ZEUS detector at HERA
\cite{zeu92}.
The kinematics of the photoproduction events was inferred from the
scattered electron registered in the electron arm of the
luminosity monitor.
Therefore the natural variables for the study of radiative effects are
leptonic variables ($x_l, y_l$ and $Q^2_l$).

The luminosity detector allowed for precise measurement of the scattered
electron in the energy range of $E'_e: 10 - 16 \gev$ and angular
range up to $6 \mrad$.
With the electron and the proton beam energies equal to
$E_e=26.6 \gev$ and  $E_p=820 \gev$, respectively,
this corresponds to  the following kinematic  limits on the leptonic
variables (3-5):
\begin{eqnarray}
\label{range}
 0.398  \le &y_l& \le 0.624     \nonumber\\
 6.9\cdot 10^{-8}   \le &Q^2_l& \le 1.5\cdot 10^{-2}    \\
 2.0\cdot 10^{-12}   \le &x_l& \le 4.4\cdot 10^{-7}    \nonumber
\end{eqnarray}
where only extreme values for $Q^2_l$ and $x_l$ were shown.
In this range  radiative corrections can be of the order of $40$ \%
depending on the $(x_l,y_l)$ point as can be seen from  table~1.
There we compared  values (in  \%)  of
\be
\label{delxy}
 \delta_{xy}=
%\frac{\frac{d^2\sigma}{dx_l dy_l}}{\frac{d^2\sigma^{Born}}{dx_l dy_l}}
%   -1
\frac{d^2\sigma^{\cal O(\alpha)}}{dx_l dy_l}
/\frac{d^2\sigma^{Born}}{dx_l dy_l} -1
\ee
obtained from TERAD91 (upper value) and HERACLES4.2 (lower value)
for the case where the radiated photon goes unobserved.
For this comparison we tried to choose the same set of input parameters.
In particular, a cut on the hadronic invariant mass $W_h >2 \gev$ was
applied in both programs and $\alpha_{QED} = const$ was used in the
calculations. Also the same structure functions (ALLM) were specified.
The agreement between these two programs  in the low $Q^2$
region is  at  the level of the statistical accuracy of the HERACLES4.2
 program ($=0.1$ \%). Only in the lowest $x$ bin
considered the difference is a little bit larger.
This table also shows that  RC are getting bigger  for
larger $x_l$ at given $y_l$ (which means also larger $Q^2_l$
values). On the other hand, they are fairly constant as a function of $y_l$
for a given $x_l$ point.

\section{Radiative events distributions}

In this section we want to discuss some characteristics of the
radiative events as produced by the program HERACLES4.2.
It must be  remembered that the notion of
"radiative events" is incomplete as  long as the minimal value
of the radiated photon energy is not specified (see e.g. \cite{kwi91}).
In HERACLES4.2 this value is determined by the program itself
and therefore radiative events distributions  (except for
the photon energy itself) do not have an immediate physical meaning.

In the following we will  look at  the  energy and the angular
distributions of the  number of    radiative events  in the region
(\ref{range}).
In general, these distributions have similar features to those
which appear in the deep inelastic region (large $Q^2$) and which were
investigated in detail in \cite{kwi91}.

In fig.1a the distribution of radiative events versus the energy
of the radiated photon is shown. The contributions from
the initial state radiation (ISR),  the final state radiation (FSR) and
the Compton part (COM) are presented separately.
The ISR and FSR distributions are
almost identical because of the small angle of the scattered electron.
In addition to peak of the soft photons at $E_\gamma = 0 \gev$ the
distribution shows also a second peak around $E_\gamma \approx 12\gev$.
All the three contributions are peaked around this value.
The  peak in the ISR (FSR) comes from the combined effect of
$1/2k\cdot p_e$ ($1/2k\cdot p_e'$) and $1/Q^2_h$ terms in the
differential cross section \cite{kwi91} ($k$ is the
4--momentum of the radiated photon).
The first term is large for collinear photon radiation from the initial
(final) state electron, i.e. small $\theta_\gamma \approx 0$.
For this situation one finds that the second term
$1/Q^2_h \sim 1/(E_e -E_\gamma)$.
The allowed maximal photon energy is then \cite{kwi91}:
\be
E_\gamma^{max}(\theta_\gamma=0)\simeq y_l\frac{1-x_l}{1-x_ly_l}E_e
\approx y_l E_e
\ee
where the last approximation comes from the very small values of $x_l$
in the considered region.
This gives for the $y_l$ range (\ref{range}) the peak at
$E_\gamma: 10.5 - 16.5 \gev$  both in the ISR and the FSR distributions.
The  peak in the Compton part  appears in the
same energy range. This  follows from the formula \cite{kwi91}:
\be
E_\gamma^C\simeq y_lE_e + x_l(1-y_l)E_p\approx y_lE_e
\ee
which fixes the energy of the emitted photon for this contribution.

The angular distribution of the radiated photon       is shown on
fig.1b together with the above mentioned three contributions
($\theta_\gamma$ is measured with
respect to the electron direction).
At $\theta_\gamma \rightarrow 0$      the initial state radiation is
 the dominant
contribution. Going towards the maximum value  of the scattered electron
($\theta_e^{max} = 6 \mrad$), the    final state radiation starts to
 determine the angular distribution.  As  in the case of the energy
distribution,  also
here the Compton contribution is much smaller than the others.
The   formula which determines the peak of the Compton part
\cite{kwi91}:
\be
 \cos \theta_\gamma^C\simeq
 \frac{y_lE_e - x_l(1-y_l)E_p}{y_lE_e+ x_l(1-y_l)E_p}
 \approx \frac{y_lE_e}{y_lE_e} = 1
\ee
confirms   that this contribution is peaked  at small angles in the
           region (\ref{range}).

\section{Cuts}

The experimental conditions at  the ZEUS detector allowed to constrain
the influence of  radiated photons in two ways:
by identifying radiated photons in the luminosity monitor and by
demanding some hadronic activity in the main calorimeter.
The photon arm of the ZEUS luminosity detector makes it possible
to register photons emitted in the angular range of
$0< \theta_\gamma < \theta_\gamma^{max} $ ($\theta_\gamma^{max}$
around $0.5-2\mrad$) and with energy  $E_\gamma$ (in average)
bigger than $0.5 \gev$.
A significant part of the hard photon radiation is produced within this
angular range \cite{akh91}.
The elastic radiation process (\ref{el}) is the dominant source
of hard photons in the luminosity counter,
exceeding the contribution of the process (\ref{inel})
by several orders of
magnitude.
Radiative corrections coming from the elastic tail are huge.
In addition, random coincidences of the non--radiative photoproduction
events and the radiative elastic events may mimic a
radiative photoproduction event of the type (\ref{inel}).
Fortunately, the careful experimental analysis allows to get rid of
the elastic tail contribution (see e.g. \cite{zeu92}).
In the following, only hard photon production coming from the inelastic
radiation process (\ref{inel}) will be investigated.

Motivated by the experimental condition of the $\sigma_{tot}$
measurement  \cite{zeu92} we specified the following  cuts   for our
     study:
\begin{description}
\item{$C_h:\ $} events with $W_h > 60\gev$ are accepted;
\item{$C_{\gamma E}:\ $} events with $E_\gamma > E_\gamma^{min}$ are
 rejected;
\item{$C_{\gamma \theta}:\ $} events with
$0 < \theta_\gamma < \theta_\gamma^{max}$ are rejected;
\item{$C_{h\gamma}:\ $} events with $W_h > 60\gev$ are accepted;
 events with $E_\gamma > E_\gamma^{min}$ and
$0 < \theta_\gamma < \theta_\gamma^{max}$ are rejected.
\end{description}
By the  "no cuts" case we will mean  the situation where
events which  fulfill the condition $W_h > 2 \gev$ are accepted.
In other cases, if not stated otherwise, $E_\gamma^{min}=0.5\gev$
and $\theta_\gamma^{max}=0.5\mrad$.

The cuts $C_{\gamma E}$ and $C_{\gamma \theta}$ are rather unphysical in
the sense that the realistic luminosity detector has restricted
angular coverage and some energy threshold for detecting photons.
Only the combined effect of these cuts
($C_\gamma=C_{\gamma E} + C_{\gamma \theta}$) corresponds to a
realistic experimental situation.  We decided, nevertheless, to analyze
separately the cuts on the radiated  photon angle and the energy in
order to see their different effect on the radiative corrections.
These cuts are implemented options in the TERAD91 program.

\subsection{Event distributions}

The effect of various cuts on the energy and angular distributions of
the radiated photon is shown in figs.2 and 3.
Fig.2a shows the influence of the $C_{\gamma \theta}$ cut on the
energy distribution for two values of the luminosity detector
angular acceptance: $\theta_\gamma^{max} = 0.5 \mrad$ (dashed line)
and  $\theta_\gamma^{max} = 2 \mrad$ (dotted line).
Only the events which remained after cuts are  shown in the figure.
For comparison, also the energy distribution without cuts is shown as
a solid line.
In fig.2b the effect of the $C_h$ cut is presented (dashed line). In
addition, the dotted line stands for
the combined cut $C_h$ and
$C_{\gamma\theta}(\theta_\gamma^{max}=0.5 \mrad)$.
It follows from these figures that the $C_h$ cut  influences mainly
events with large energy, reducing the second peak in the
energy distribution. On the other hand, the cut on the angle of
the radiated photon reduces the number of radiative events in the
whole energy range.
The arrow in fig.2b corresponds to the $E_\gamma^{min}=0.5 \gev$
value; all the events above this value are rejected by the
$C_{\gamma E}$ cut.

The angular distribution of  radiated photon in  the presence
of various cuts is shown in fig.3.
As before, the solid lines stand for the "no cuts" situation.
The number of events after the $C_{\gamma E}$ cut with
$E_\gamma^{min}=0.5 \gev$ (dashed line) and
$E_\gamma^{min}=3   \gev$ (dotted line) is presented in fig.3a.
For comparison, in fig.3b,  the angular distribution after
$C_h$ is shown (dashed line) together with
the combined cut:
 $C_h+C_{\gamma E}(E_\gamma^{min} = 0.5\gev)$ as dotted line. It is seen
that the $C_{\gamma E}$ cut is more
restrictive   than $C_h$ in the whole range of
the analyzed $\theta_\gamma$.
In particular, the dashed line  in fig.3a and the dotted line in
fig.3b are exactly the same.
Finally, the arrow in fig.3b  specifies the position of the
photon  angle $\theta_\gamma^{max}= 0.5 \mrad$. All the events with
smaller $\theta_\gamma$ are rejected by the $C_{\gamma \theta}$ cut.

\subsection{Radiative corrections}

The influence of the above cuts on the size of     RC
can be very different. In the following we will analyze
this influence in more details.

First we studied   radiative corrections to the double
differential cross section as a function of $x_l$
in the middle of the considered $y_l$ range: $y_l=0.49$ (fig.4,5).
These results were obtained from the TERAD91 program.
In the presence of a cut, radiative corrections were calculated
from the remaining number of events.

Fig.4 shows how the radiative corrections are affected by cuts
on the energy ($C_{\gamma E}$) and the angle ($C_{\gamma \theta}$) of
the radiated photon.
As it was already pointed out before,
 RC without any cuts (solid line) can be quite large.
The $C_{\gamma E}$ cut (dotted lines) strongly reduces their
value--at large $x_l$ even by $30$ \%.
In case of $E_\gamma^{min}=3 \gev$ (upper dotted line)
$\delta_{xy}$ is almost vanishing in the considered $y_l$ bin.
 For lower $E_\gamma^{min}=0.5\gev$ (lower dotted line)
its value is $\delta_{xy}=-12\%$ at large $x_l$.
One should, however,  remember that this cut cannot be realized
           experimentally (see discussion above).
The  cut imposed on the angle of the radiated photon,  $C_{\gamma
\theta}$, makes $\delta_{xy}$  behave in a more
complicated way (dashed lines on fig.4).
The size of radiative corrections in this situation strongly
depends on the angular acceptance of the luminosity detector.
For example, the upper dashed line represents
$\theta_\gamma^{max}=0.5\mrad$ while the lower dashed
line  $\theta_\gamma^{max}=2\mrad$.
Also this cut is not feasible experimentally.
Only the combination of the above two cuts
($C_\gamma=C_{\gamma \theta}+C_{\gamma E}$) can be applied in a
realistic experimental situation. The dash-dotted lines show
the influence of this cut on  the RC.
The upper dash-dotted line corresponds to
$E_\gamma^{min}=0.5\gev,\ \theta_\gamma^{max}=0.5\mrad$
 and the lower dash-dotted line -- to
$E_\gamma^{min}=0.5\gev,\ \theta_\gamma^{max}=2\mrad$.
It is  seen that in the region of low $x_l$ (low $Q^2_l$)
the cut on radiated photon energy dominates while at larger $x_l$
the angular cut determines the size of $\delta_{xy}$.
In conclusion, the  combined cut on the photon energy  and angle
reduces significantly the size of radiative corrections,
especially at lower $x_l$. However, at the upper part of  the
$x_l$ ($Q^2_l$) range these corrections are still at the level of
$20$ \%.

In fig.5 we consider also effects coming from the
cut on the hadronic invariant mass, $W_h$.
The  solid line represents again the "no cuts"
situation. The dotted line corresponds to the
$C_h$ cut. This cut alone reduces substantially the
radiative corrections from $40$\% to $10$\% at large $x_l$.
The addition  of the  constraints on the radiated
photon energy and angle ($C_{h\gamma}$ cut) makes     RC
almost negligible.
The dash-dotted line represents  the   $C_{h\gamma}$ cut
with $E_\gamma^{min}=0.5\gev,\ \theta_\gamma^{max}=0.5\mrad$,
the upper dashed line:
     $E_\gamma^{min}=3\gev,\ \theta_\gamma^{max}=0.5\mrad$
and the lower dashed line:
     $E_\gamma^{min}=0.5\gev,\ \theta_\gamma^{max}=2\mrad$.
Comparison of figs.4 with 5 makes it clear that the addition of
the  hadronic cut to the constraints of the radiated photon phase space
allows to obtain small
radiative corrections in the whole $x_l$ ($Q^2_l$) range of the
considered $y_l$ bin.
While the numerical values of $\delta_{xy}$ depend on the specific
$y_l$ bin, the qualitative  behaviour of the radiative corrections in
other $y_l$ bins is similar to that  found in $y_l=0.49$ bin.

Measurement of the scattered electron permits, in principle, to
study the photoproduction cross section in a given $y$ bin.
It is worth, therefore, to look at the size of radiative corrections
to the cross section (\ref{xrad}) integrated over the allowed
$x_l$ range:
\be
 \delta_{y}=
%\frac{\frac{d\sigma^{\cal O(\alpha)}}{dy_l}}
%{\frac{d\sigma^{Born}}{dy_l}} -1
\frac{d\sigma^{\cal O(\alpha)}}{dy_l}/\frac{d\sigma^{Born}}{dy_l} -1.
\ee
Fig.6 shows $\delta_y$ as a function of $y_l$ in  the presence of
various cuts calculated by   TERAD91 and HERACLES4.2.
The lines correspond to the results of TERAD91 while
the points are calculated from HERACLES4.2.
The solid line  represents the "no cuts" situation, the dotted
 one -- after $C_h$ is imposed, the dashed line -- with the $C_\gamma$
with
$E_\gamma^{min}=0.5\gev, \theta_\gamma^{max}=0.5\mrad$ and the
dash-dotted line, the combination of the above two cuts: $C_{h\gamma}$.
The first observation is that the  radiative corrections to the
once integrated cross section are in general smaller than those
for the double differential cross section (\ref{delxy}) and are
almost flat over the whole  $y_l$ range.
These features can be understood in the following way:
the size of $\delta_{xy}$ is growing with $x_l$ for
a given $y_l$ (i.e. also growing with $Q^2_l$).
However,   the cross section (\ref{xrad}) is strongly peaked
at small $Q^2_l$ values
because of the propagator factor. In other words, it
     is getting smaller with growing $Q^2_l$.
As a result, the   $x_l$ bins with lower radiative corrections
are giving bigger contributions to the integral over $x_l$.
With the $C_{h\gamma}$ cut imposed  the value of the radiative
corrections
 to the once integrated cross section amounts to around $-1 \%$.
This fact permits to interpret the measured cross section
to a very good approximation as the Born one.
The  figure also shows very good agreement between the TERAD91
and the HERACLES4.2 programs.

As a last step in our analysis we calculated  RC to the
total photoproduction cross section in the region defined by
(\ref{range}).
The results from both programs are presented
in table~2. It is seen that radiative corrections to  $\sigma_{tot}$
can be substantially reduced by applying experimentally
feasible  cuts. In particular, after imposing the $C_{h\gamma}$ cut
with $E_\gamma^{min}=0.5\gev$ and $\theta_\gamma^{max}=0.5\mrad$,
the radiative corrections to the total photoproduction cross section
are almost negligible.

In the end we want to comment about the
dependence of our results on the structure functions.
It was already pointed out in sec.2   that the size of  RC
(hard photon contribution) depends on the structure functions used
in the calculations.  This dependence was clearly observed
in  the deep inelastic region (see for example \cite{kwi91}) and was
caused mainly by  uncertainties of extrapolation of
the structure functions to the unmeasured part of the region.
In order to estimate the influence of the structure functions
on the radiative corrections in the photoproduction region
one needs to have at least two different parametrizations
of structure functions which  both agree with existing
photoproduction  data and which are also compatible  with
the recent data on the deep inelastic structure functions.
At the moment only the ALLM parametrization fulfills
these conditions.
However, one would expect a very small  dependence
of  this kind in the photoproduction region, as
the existing low $Q^2$  data constrain possible parametrizations
of the structure functions in this region quite well
\footnote{Conversations with D.Bardin about this point are
acknowledged.}.

%We analyzed this effect by using instead of
%the ALLM prescription the Duke and Owens parametrization of
%parton distributions \cite{do84} modified in the low $Q^2$ range
%$(Q^2_h\le 4 \gev^2)$ by a multiplicative factor \cite{spi92}:
%\be
%   (1-\exp(-a^2Q^2_h)),\qquad a^2=3.37 \gev^{-02}.
%\ee
%The effect of different structure functions (ALLM vs DO) can change the
%results for $\delta_{xy}$ by  $15-19\%$. This    comes from the  fact
%that both low $Q^2$ behaviour and small $x$ dependence of the structure
%functions  influences calculation of  QED  leptonic corrections
%(through the second term in (\ref{xrad})).
\vspace{1cm}

\section{Conclusions}

We have analyzed the  behaviour of radiative corrections in the region of
low $Q^2$ at the HERA $ep$ collider.
The study of various cuts suggests that  the  $C_{h\gamma}$ cut is
the most favourable.
It  requires
observation  of  some hadronic activity in the
main part of the calorimeter and  detection of the radiated
photon in the luminosity monitor. This cut allows to reduce
the size of radiative corrections for the
once integrated cross section (over $x$) and for the total
 photoproduction cross section down to the level of  $-1\%$.
The results  obtained with the help of two different
programs for calculating radiative corrections,
 TERAD91 and HERACLES4.2, agree very well in the considered region.

\vspace{1cm}

\noindent {\Large\bf Acknowledgments}

We thank  A.Levy for suggesting us this study and for
encouraging discussions.
We would also like to thank D.Bardin and H.Spiesberger for
comments concerning radiative corrections  in general and their
 programs in  particular.
We are indebted to  G.Wolf for a careful reading  the manuscript
and remarks.
This work has been supported    partially  by the  KBN PB 1071/2/91
grant.

%--------- REFERENCES -------------

\newpage

\noindent {\Large \bf Figure captions}
\begin{enumerate}

\item %1
 The energy (fig.1a) and angular (fig.1b) distribution of
 the number of radiated photon events.
 The dashed line corresponds to the initial state radiation,
 the dotted line -- to the final state radiation and
 the dash--dotted line refers to the Compton part.
 The solid line denotes the sum of the above three contributions.

\item %2
 The energy distribution of the radiative events in the
presence of various cuts. In fig.2a   the
photon phase space constraints are presented:
the solid line -- no cuts;              $C_{\gamma \theta}$ cut with
$\theta_\gamma^{max}=0.5\mrad$ -- the dashed line and with
$\theta_\gamma^{max}=2\mrad$ -- the dotted line.
Fig.2b presents the effect of the hadronic cuts: the solid line -- no cuts;
the dashed line -- $C_h$ cut and the dotted line -- $C_{h\gamma\theta}$ cut
with $\theta_\gamma^{max}=0.5\mrad$.

\item %3
 The angular distribution of the radiative events in the
presence of various cuts. In fig.3a   the
photon phase space constraints are presented:
 the solid line -- no cuts;              $C_{\gamma E}$ cut with
$E_\gamma^{min}=0.5\gev$ -- the dashed line and with
$E_\gamma^{min}=3\gev$ -- the dotted line.
Fig.3b presents the effects of the hadronic cuts: the solid line -- no cuts;
the dashed line -- $C_h$ cut and the dotted line -- $C_{h\gamma E}$ cut with
$E_\gamma^{min}=0.5\gev$.

\item %4
Effects of the photonic   cuts on the size of the
radiative corrections in the $y_l=0.49$ bin:
the solid line -- no cuts;
the dotted lines -- $C_{\gamma E}$ cut with $E_\gamma^{min}=3\gev$
 (upper line) and $E_\gamma^{min}=0.5\gev$ (lower line);
the dashed lines -- $C_{\gamma \theta}$ cut with
$\theta_\gamma^{max}=0.5\mrad$ (upper line)
$\theta_\gamma^{max}=2\mrad$ (lower line);
the dash-dotted lines -- combined $C_\gamma=C_{\gamma E}+C_{\gamma \theta}$ cut
with $E_\gamma^{min}=0.5\gev, \theta_\gamma^{max}=0.5\mrad$ (upper line)
and $E_\gamma^{min}=0.5\gev, \theta_\gamma^{max}=2\mrad$ (lower line).

\item %5
Effects of the combined (photonic and hadronic) cut $C_{h\gamma}$
on the size of the radiative corrections in the $y_l=0.49$ bin:
the dash-dotted line -- with
$E_\gamma^{min}=0.5\gev, \theta_\gamma^{max}=0.5\mrad$;
the upper dashed line -- with
$E_\gamma^{min}=3\gev, \theta_\gamma^{max}=0.5\mrad$;
the lower dashed line -- with $E_\gamma^{min}=0.5\gev,
\theta_\gamma^{max}=0.5\mrad$.
For comparison also hadronic cuts are shown:
the  solid line corresponds to the "no cuts"
situation (i.e. events with $W_h>2 \gev$ are accepted)  and the
dotted line represents the
$C_h$ cut (i.e. events with $W_h>60 \gev$
are accepted).

\item %6
The size of radiative corrections to the single integrated cross
section in the presence of various cuts.
The lines correspond to results of TERAD91 while
the points are calculated from HERACLES4.2.
The solid line  represents the "no cuts" situation, the dotted
 one -- $C_h$ cut, the dashed line -- $C_\gamma$ cut
with $E_\gamma^{min}=0.5\gev, \theta_\gamma^{max}=0.5\mrad$ and the
dash-dotted line -- the combination of the above two cuts($C_{h\gamma}$).
\end{enumerate}

\newpage
\begin{table} \label{tab1}
\begin{center}
\begin{tabular}{|l|r|r|r|r|r|c|}
\hline
\hline
     & & & & & &\\
$~~x_l~~\backslash~~~~y_l$ & $0.40$ & $0.45$ & $0.50$ & $0.55$ & $0.60$
& program \\
     & & & & & &\\
\hline
\hline
     & & & & & &\\
$10^{-11}$ & $1.6$ & $ 1.9$ & $ 2.2$ & $ 2.5$ & $2.9$ & TERAD91\\
           & $1.2$ & $ 1.4$ & $ 1.6$ & $ 1.7$ & $1.8$ & HERACLES42\\
     & & & & & &\\
\hline
     & & & & & &\\
$10^{-10}$ & $6.7$ & $ 7.3$ & $ 7.9$ & $ 8.4$ & $9.0$ & TERAD91\\
           & $6.6$ & $ 7.2$ & $ 7.8$ & $ 8.3$ & $8.8$ & HERACLES42\\
     & & & & & &\\
\hline
     & & & & & &\\
$10^{-9}$  & $14.7$ & $15.5$ & $16.3$ & $17.0$ & $17.6$ & TERAD91\\
           & $14.5$ & $15.3$ & $16.0$ & $16.8$ & $17.5$ & HERACLES42\\
     & & & & & &\\
\hline
     & & & & & &\\
$10^{-8}$  & $23.2$ & $24.2$ & $25.1$ & $25.9$ & $26.7$ & TERAD91\\
           & $23.3$ & $24.2$ & $25.2$ & $26.0$ & $26.6$ & HERACLES42\\
     & & & & & &\\
\hline
     & & & & & &\\
$10^{-7}$  & $31.7$ & $32.9$ & $33.9$ & $34.8$ & $35.6$ & TERAD91\\
           & $31.5$ & $33.0$ & $34.2$ & $35.2$ & $36.0$ & HERACLES42\\
     & & & & & &\\
\hline
\hline
\end{tabular}
\caption{{\it Results for the NC QED leptonic  corrections
$\delta_{xy}$ (in \%) in various points of leptonic
($x_l,y_l$) variables ("no cuts" situation).}}
\end{center}
\end{table}

\begin{table} \label{tab2}
\begin{center}
\begin{tabular}{|r|r|r|r|c|}
\hline
\hline
      & & & &\\
  $C_0$  & $C_h$   & $C_\gamma$ & $C_{h\gamma}$ & program \\
      & & & &\\
\hline
\hline
      & & & &\\
 $17.9$ & $4.6$ & $4.8$ & $-1.2$ & TERAD91\\
 $18.6$ & $5.2$ & $5.6$ & $-0.8$ & HERACLES42\\
      & & & &\\
\hline
\hline
\end{tabular}
\caption{{\it Radiative corrections $\delta$ (in \%) to the
total photoproduction cross
section with various cuts explained in the text. }}
\end{center}
\end{table}

\end{document}